\documentclass[10pt,a4paper]{article}
\usepackage{amsmath}
\usepackage{amsthm}
\usepackage{fancybox}
\usepackage{bbm}
\usepackage{pdflscape}
\usepackage{algorithm}
\usepackage{algpseudocode}
\usepackage{caption}
\usepackage{subcaption}
\usepackage{placeins}
\usepackage{array}
\usepackage{multirow}
\usepackage{marvosym}
\usepackage{longtable}
\usepackage{natbib}
\usepackage{textcomp}
\usepackage{rotating}
\usepackage{enumitem}
\usepackage{pifont}
\usepackage{url}
\usepackage{xspace}
\usepackage{multirow}

\newcommand{\E}{\mathbbm{E}}

\newcommand{\M}{\ensuremath{\mathcal{M}}}

\newcommand\LGD{\ensuremath{{\rm LGD}}} 
\newcommand\lgd{\LGD}

\newcommand\TF{\ensuremath{{\rm termsheet\ flows}}}
\newcommand\CSA{\ensuremath{{\rm CSA\ flows}}}
\newcommand\IM{\ensuremath{{\rm initial\ margin}}}
\newcommand\SF{\ensuremath{{\rm settlement\ flows }}}

\newcommand\SA{{SwapAgent\textsuperscript{\textregistered}}\xspace}
\newcommand\ISDA{{ISDA\textsuperscript{\textregistered}}\xspace}

\newcommand\CDF{\ensuremath{{\rm CDF}}}
\newcommand\PFE{\ensuremath{{\rm PFE}}}
\newcommand\PFL{\ensuremath{{\rm PFL}}}
\newcommand\aPFL{\ensuremath{{\rm aPFL}}}
\newcommand\paPFL{\ensuremath{{\rm paPFL}}}
\newtheorem{define}{Definition}

\begin{document}
\author{Chris Kenyon, Mourad Berrahoui and Benjamin Poncet\footnote{Contacts: chris.kenyon@mufgsecurities.com, mourad.berrahoui@lloydsbanking.com, benjamin.poncet@lloydsbanking.com.   
\vskip3mm
The views expressed in this presentation are the personal views of the authors and do not necessarily reflect the views or policies of current or previous employers. Not guaranteed fit for any purpose.  Use at your own risk.
\vskip3mm This paper is a personal view and does not represent the views of MUFG Securities EMEA plc (“MUSE”).  This presentation is not advice.  Certain information contained in this presentation has been obtained or derived from third party sources and such information is believed to be correct and reliable but has not been independently verified.  Furthermore the information may not be current due to, among other things, changes in the financial markets or economic environment.  No obligation is accepted to update any such information contained in this presentation.   MUSE shall not be liable in any manner whatsoever for any consequences or loss (including but not limited to any direct, indirect or consequential loss, loss of profits and damages) arising from any reliance on or usage of this presentation and accepts no legal responsibility to any party who directly or indirectly receives this material.
}}
\title{Counterparty Trading Limits Revisited:\\ CSAs, IM, \SA\\From PFE to PFL}
\date{23 November 2018\vskip5mm Version 2.7, {\it to appear in Risk}}

\maketitle

\begin{abstract}
Potential Future Exposure (PFE) for counterparty trading limits is challenged by new market developments, notably widespread regulatory Initial Margin, and netting of trade and collateral flows --- but PFE already has many issues, e.g. comparability across counterparties. We introduce Potential Future Loss (PFL) which combines expected shortfall (ES) and loss given default (LGD) as a replacement for PFE and provide extensions to cover the main issues with PFE. 
\end{abstract}

\section{Introduction}

The utility of Potential Future Exposure (PFE) for counterparty trading limits is being challenged by new market developments, notably widespread regulatory Initial Margin \citep{bcbs2015im}, and netting of trade and collateral flows (e.g. via \SA, \cite{lch2018swapagent}). However PFE already has challenges: counterparty trading limits are not comparable across counterparties because of varying recovery rates, and because of different loss distributions above the reference quantile for PFE. In addition, PFE limits are typically changed when a collateral agreement (Collateral Support Annex, or CSA, of the \ISDA\footnote{International Swaps and Derivatives Association master agreement.}) is put in place. That is, the effects of the change in loss distribution and any potential change in recovery are included ad hoc. Furthermore, trading limits with the same counterparty but at different seniorities are not fungible because Credit Officers take into account the differing recoveries for different seniorities. In addition overlaps with credit mitigation and CVA are not included in PFE. Thus the typical counterparty limit metric, PFE, as a high quantile (95\% to 99\%) of future exposures, needs updating. We introduce Potential Future Loss (PFL) which combines expected shortfall (ES) and loss given default (LGD). With two additional variants Adjusted PFL (aPFL) and Protected Adjusted PFL (paPFL) these deal with pre-existing challenges and the new ones. 
PFE is generally defined as:
\begin{define}[$\PFE_\M$(t,q)] Potential Future Exposure at time $t$ in the future for quantile $q$ under measure $\M$ is
\[
\PFE_\M(t,q) := \CDF_\M^{-1}(q)\left(\max(V(\Pi,t,\delta_B,\delta_C), 0)\vphantom{\int}\right) 
\]
\end{define}
\noindent
where $V(.)$ is the value of the portfolio $\Pi$ in the netting set of interest, considering cashflow timing assumptions $\delta_B,\delta_C$ on termsheet and collateral and/or settlement flows, conditional on default (notation is given in Table \ref{t:notation}). The cashflow types timing expands on \cite{andersen2017rethinking} to include initial margin flow timing and settlement flow timings. Thus our definition of \PFE\ includes the effects of collateral, settlement, and initial margin (cleared or uncleared).

\begin{table}
\begin{tabular}{lp{9cm}}
\bf notation & \bf description \\ \hline
& \\
$\CDF^{-1}_\M(q)(...)$ & inverse Cumulative Distribution Function of $(...)$ for the quantile $q$ under measure $\M$ \\
$V(\Pi,t,\delta_B,\delta_C)$ & value of the portfolio $\Pi$ conditional on default\\
$\delta_*,\quad*\in\{B,C\}$ & vector with components \{\TF, \CSA, \SF, \IM\} containing the timing on last cashflow of each type by each counterparty * prior to default\\
$X$ & incurred CVA; note that this is a constant and has no profile\\
$Y(t)$ & profile of credit protection\\
\end{tabular}
\caption{Notation}
\label{t:notation}
\end{table}

The measure \M\ is often chosen as the inverse-$T$-Forward measure which is defined as the risk-neutral value (which is measure-independent) inverse-discounted by an observed discount curve (which implicitly selects the $T$-Forward measure). Inverse-discount means divide by the discount factor. Historical volatilities may be used in place of market-implied volatilities. Choosing \M\ is out of scope but discussed elsewhere \citep{kenyon2015pfe,stein2015measures}.

\section{Challenges to PFE}

The effect of widespread regulatory Initial Margin \citep{bcbs2015im} on PFE was the initial motivation for our reassessment of PFE, but this reassessment reveals that PFE has existing issues as shown in Table \ref{t:issues}. We will now comment briefly on each of the issues before introducing PFL which addresses all the issues with PFE in Table \ref{t:issues}. It may seem ambitious to attempt to solve so many issues at once but in fact there are only two driving factors (often intertwined) for the issues with PFE: recovery rates and loss distribution effects. These lead naturally to our proposal for PFL as expected shortfall times loss given default.
\begin{table}
\begin{tabular}{llp{7cm}p{3cm}}
\bf & \bf Significance & \bf Issue with PFE & \bf Main Source\\ \hline
\multirow{7}{*}{Now}
&Major &Lack of comparability across counterparties in different sectors & Recoveries\\
&Major &Lack of comparability between counterparties with and without collateral & Distribution shapes\\
&Major &Lack of consistency with credit mitigation & Credit mitigation ignored \\
&Major &Insensitivity to exposure portfolio/distribution effects & Distribution shapes\\
&Medium &Insensitivitiy to existing credit losses, i.e. CVA, that has already gone through PnL& Incurred CVA ignored\\
&Medium &Lack of comparability before and after when collateralization is introduced & Distribution shapes\\
&Medium &Lack of comparability within a counterparty for netting sets of different seniorities& Recoveries\\ \hline \hline
\multirow{2}{*}{New}
&Medium &Widespread regulatory Initial Margin. Phased in from 2016-2020 & Distribution shapes\\
&Medium &Netting of collateral mark-to-market flows and trade termsheet cash flows, e.g. \citep{lch2018swapagent} which started in 2017. & Distribution shapes\\ \hline
\end{tabular}
\caption{Existing and new issues with PFE. Issues are things that significantly reduce usefullness or accuracy.}
\label{t:issues}
\end{table}

\subsection{Lack of comparability across counterparties in different sectors}

PFE limits for a counterparty in one sector is not comparable with PFE for a counterparty in a different sector because the expected recovery after default can be wildly different, \cite{jankowitsch2014determinants} find 16\%\ to 60\%\ across major sectors. Even within a sector the PFE limits may not be comparable for different subsectors, consider recoveries on Savings and Loan (median one percent) vs Credit and Financing (median 65\%).   Thus it is difficult to assess how the risk appetite of the bank is being put into practice. By itself PFE does not indicate the risk appetite of the bank thus impeding efficient risk management.

\subsection{Lack of comparability between counterparties with and without collateral\\ Lack of comparability before and after when collateralization is introduced}

\begin{figure}[htbp]
\centering
\includegraphics[width=0.45\textwidth,clip,trim=0 0 0 0]{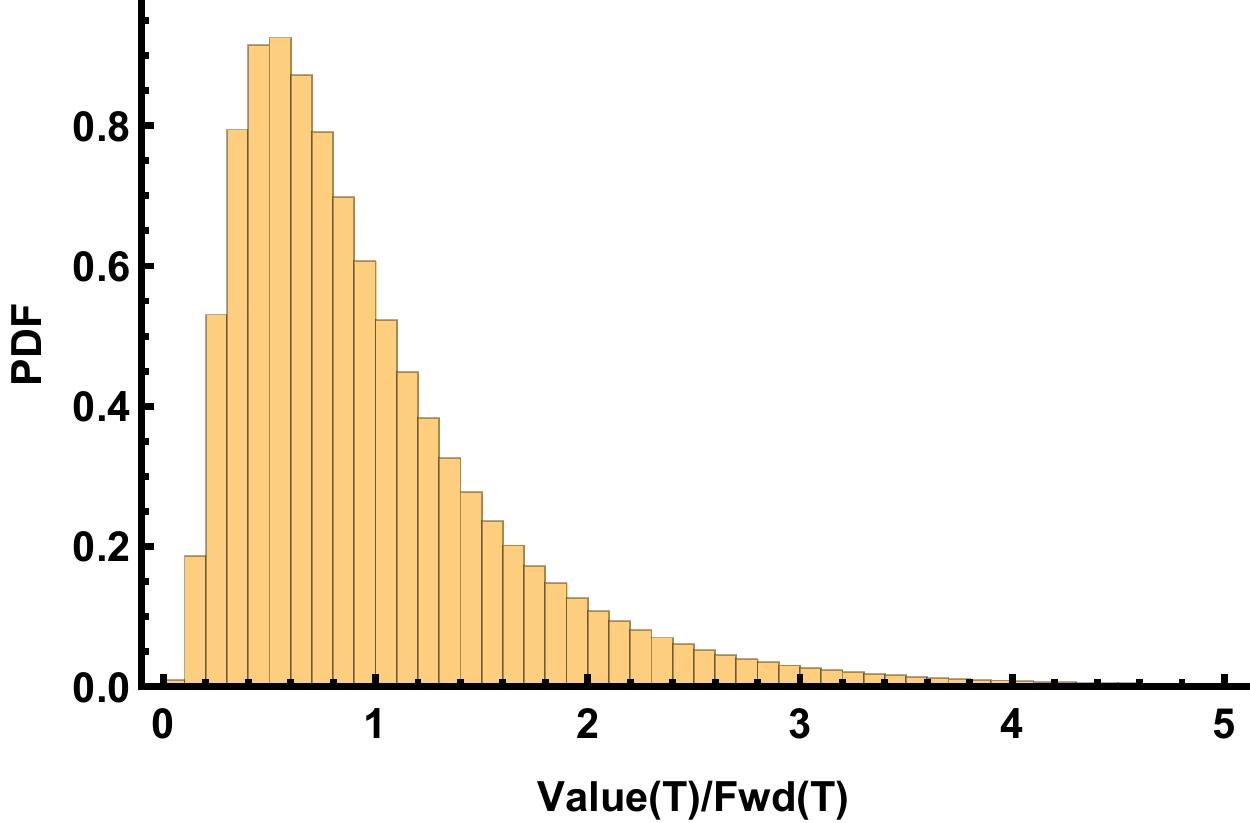}
\includegraphics[width=0.45\textwidth,clip,trim=0 0 0 0]{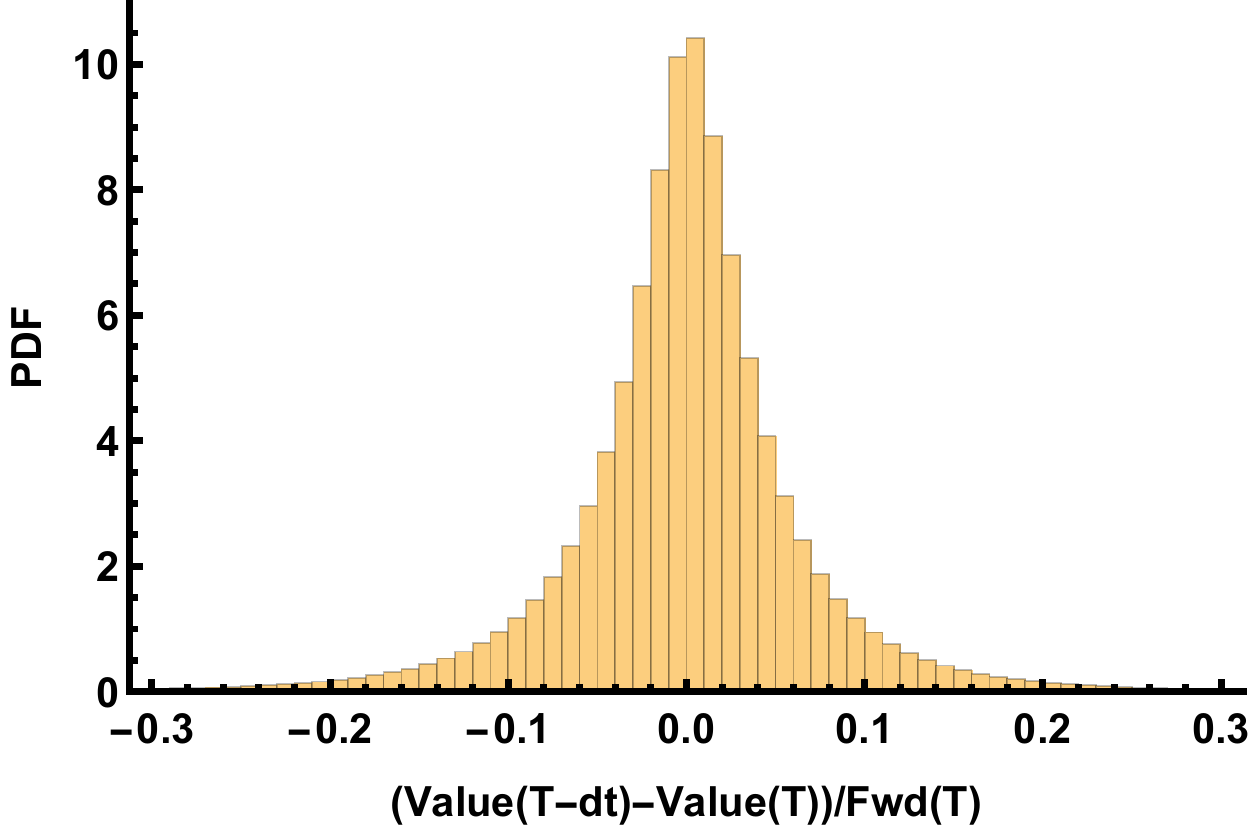}
\caption{Considering one time-point on the exposure profile where there is a LogNormal probability distribution function (PDF) of exposure LEFT, the collateralized (Calendar Spread) exposure changes to the PDF at RIGHT (very close to a Student-t with 2 degrees of freedom). Setup details: Geometric Brownian motion, drift 1\%, volatility 20\%, MPOR 2 weeks (for collateralized).}
\label{f:problemCollateral}
\end{figure}

Collateralization has two effects w.r.t. uncollateralized exposure: change in loss distribution; and change in recovery rate. PFE cannot capture either of these effects, and we examine the numerical significance later. This means that the PFE limits for collateralized counterparties cannot be compared to those without collateral. Nor can limits before collateralization be compared to limits after collateralization. This makes effective risk management more difficult.

With collateralization the exposure distribution changes from a strip of European Call options (uncollateralized) to a strip of Calendar Spread Call options. Figure \ref{f:problemCollateral} illustrates the change. In addition, the effect of the distribution changes will be portfolio dependent.

When a collateralized counterparty defaults this is typically because it has debts. Some of these will be via collateralized counterparties. The default mechanism is often that it cannot raise liquidity to pay collateral calls. In short, assets (or financialized assets) pledged as collateral are not available to creditors. Thus, all other things being equal, we can expect lower recoveries from collateralized counterparties than uncollateralized, all other things being equal.

\subsection{Lack of consistency with credit mitigation}

If a desk wishes to trade with a counterparty and the PFE limit is full, it may buy credit mitigation (e.g. a CDS) and then d with the Credit Officer about how much capacity this creates for trading. This is inefficient both from a time point of view and from potential variability between Credit Officers' applications of guidelines. There is less credit risk but PFE and PFE limits have no way to automatically include such credit mitigation.  We assume that the PFE system automatically includes Independent Amounts, detailed collateral terms, etc.  Of course only a CDS that references the actual counterparty automatically reduces exposure on default. 

\subsection{Insensitivity to exposure portfolio/distribution effects}

Since PFE(q) is an exposure quantile it is insensitive to any changes of exposure distribution above $q$. Thus there can be arbitrary changes in exposure --- provided they are 1-in-20 {\it at any time}, for $q$=95\%, say. This means that Credit Officers have to factor in these possibilities by hand when setting PFE limits.

The distribution-insensitivity of PFE is worse than it appears because the tail of the portfolio-dependence of the exposure distribution. That means that a change in the trading pattern of a counterparty can change the exposure above $q$ and this will not show up. This risk-insensitivity of PFE for relatively common (1-in-20 for 95\%\ PFE) events is undesirable. Suppose now that Credit Officers change their $q$ from 95\%\ to 99\%. This has two effects: firstly Credit Officers and Relationship manager have to re-calibrate their risk understanding; and secondly the PFE limits have to be increased for all counterparties. Even if this is done, there is now an in-sensitivity to 1-in-100 events: and two or three can be expected each year, per counterparty.

\subsection{Insensitivity to existing credit losses, i.e. CVA, that has already gone through PnL}

PFE is insensitive to CVA losses that have already been incurred. Basel III deducts incurred CVA from exposure at default in capital calculations on the grounds that this loss has already gone through PnL (\cite{bcbs2012faq}, Section 2d). It is not reasonable that a credit limit should ignore credit losses, nor is it reasonable that a metric used for credit control should ignore credit losses. However, this is the case for PFE and PFE limits unless the limits are manually changed. This is poor credit risk management.

\subsection{Lack of comparability within a counterparty for netting sets of different seniorities}

If there are multiple netting sets with the same counterparty at different seniorities then this is a challenge to PFE. Typically there will be separate counterparty trading limits against each netting set. However, the risk is to the counterparty not the netting sets so this is an issue. In addition it is generally not possible, nor desired, to move limit capacity from one netting set to another with a different seniority 1-for-1. 

Different recovery rates is the main reason that Credit Officers have different appetites for PFE for netting sets at different seniorities (e.g. 
\cite{jankowitsch2014determinants} finds median recovery for unsecured at 42\%\ and for subordinated at 5\%
). PFE does not take this into account, but the Credit Officers do --- hence limit management inefficiency and lack of comparability even within a single counterparty and lack of fungibility.

\subsection{Widespread regulatory Initial Margin}

\begin{figure}[htbp]
\centering
\includegraphics[width=0.45\textwidth,clip,trim=0 0 0 0]{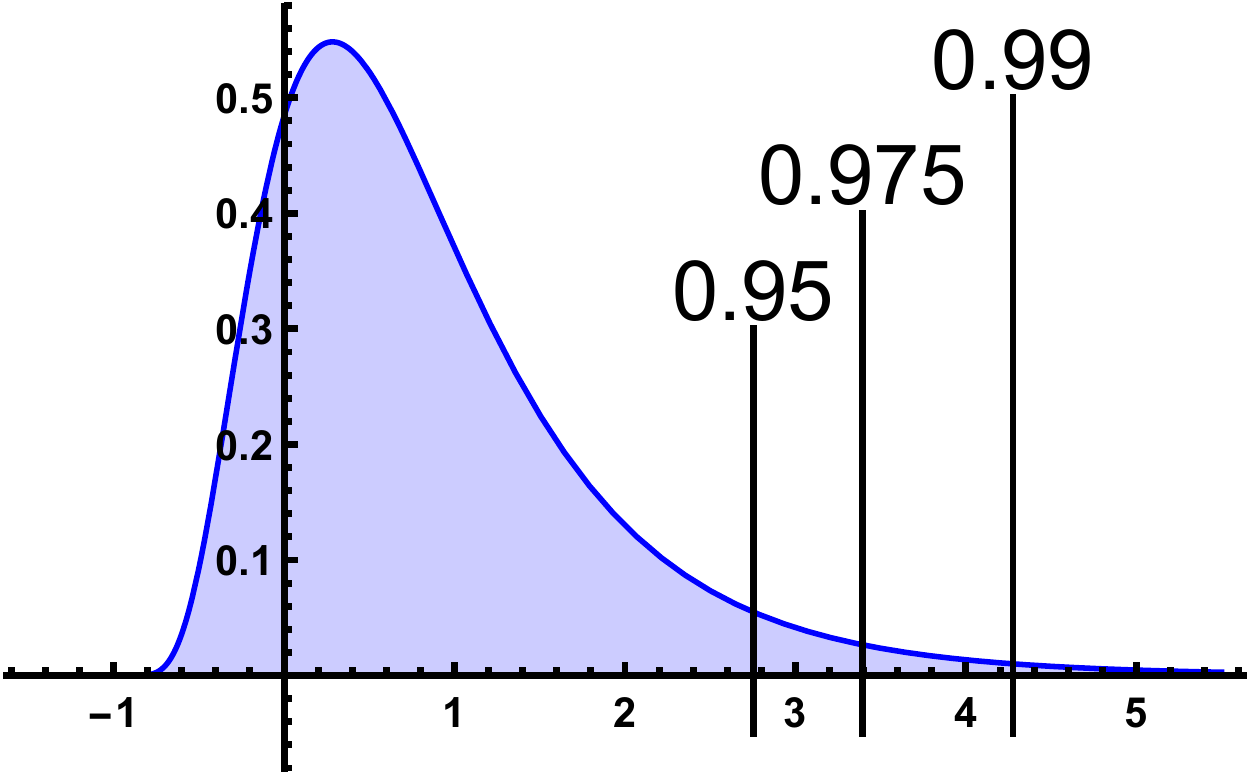}
\includegraphics[width=0.45\textwidth,clip,trim=0 0 0 0]{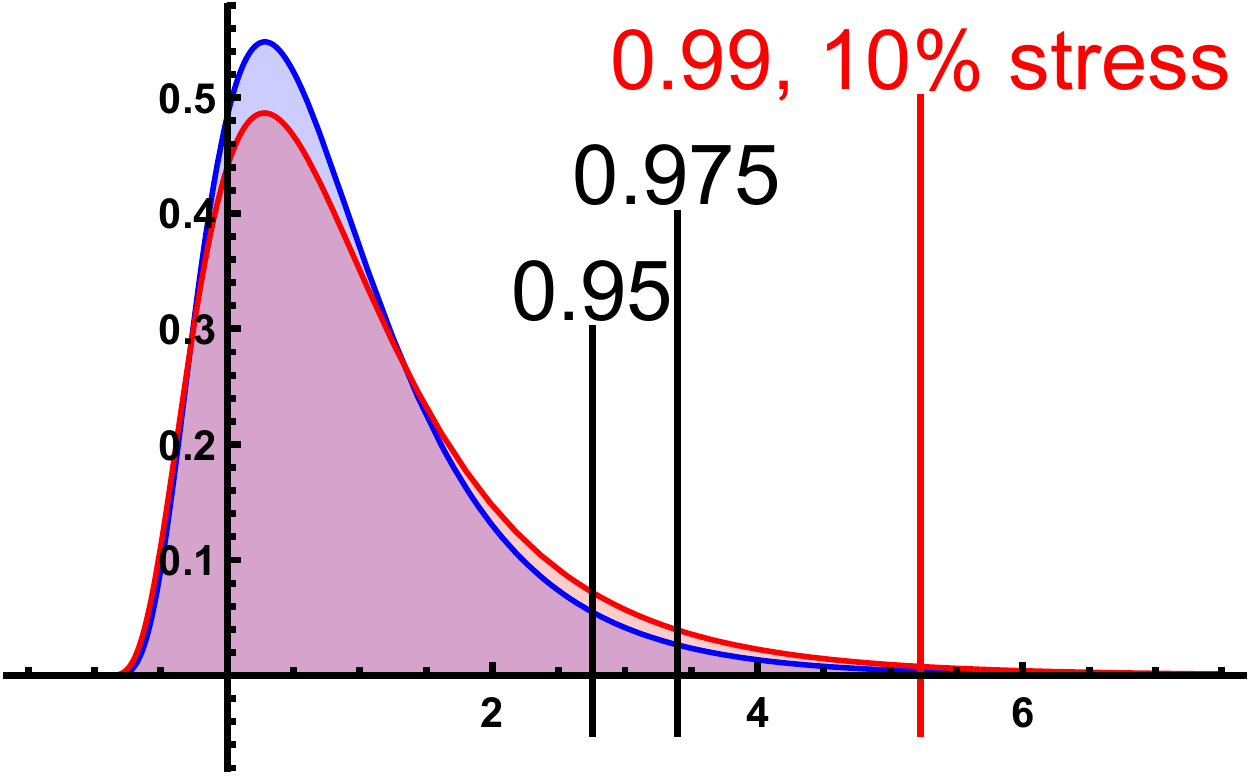}
\includegraphics[width=0.45\textwidth,clip,trim=0 0 0 0]{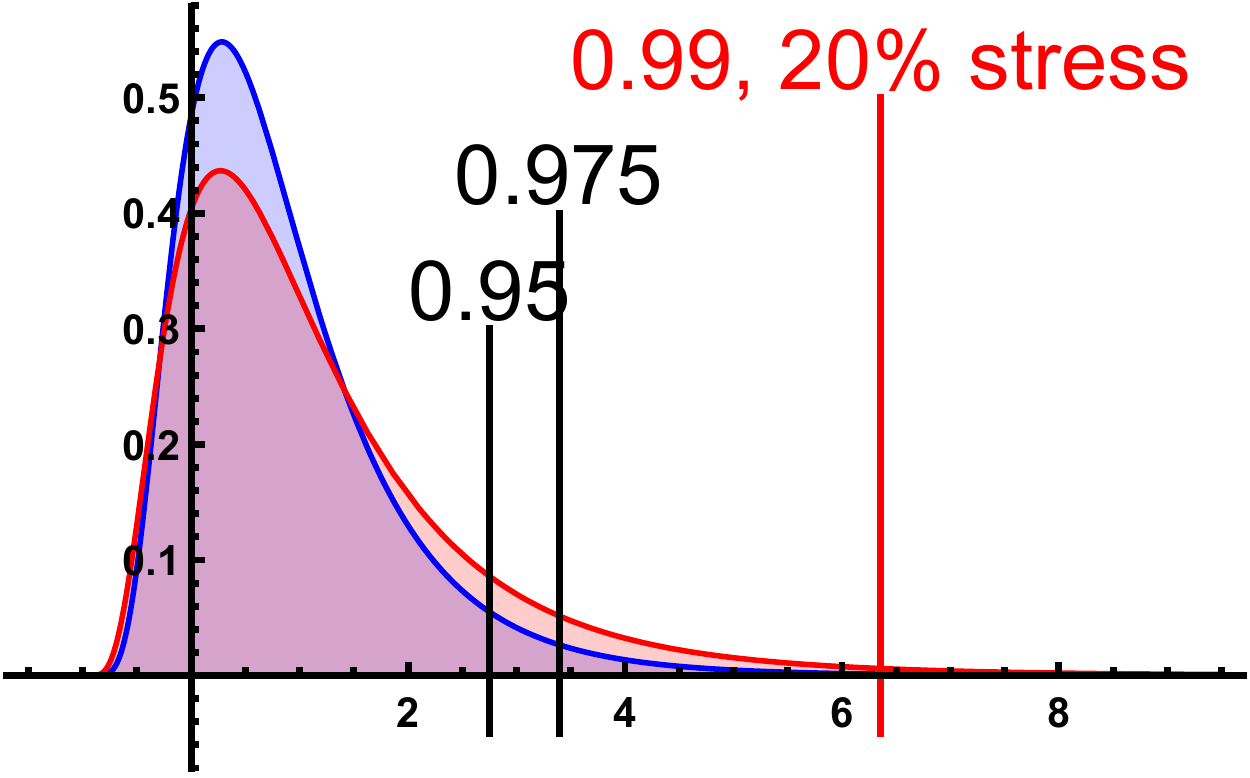}
\caption{The IM challenge to PFE. Shifted LogNormal exposure with quantiles, including the 99\%\ level defined in regulations for Initial Margin (TOP LEFT). Comparison with 99\% quantile given a 10\%\ volatility increase (stress), TOP RIGHT; and a 20\%\ volatility increase (BOTTOM). Given IM defined as 99\% 10-day one-sided exposure, PFE appears identically zero in all three cases (ignoring the issue of return of collateral spikes in exposure covered in the next section).}
\label{f:problemIM}
\end{figure}

One regulatory IM definition is as a 10-day, one-sided 99\%\ exposure, calibrated to a period of stress. Alternatively, a schedule-based method which uses a lookup table based on notional and maturity can be applied. The schedule-based method makes no allowance for netting so most large traders will use the exposure method. Note that the margin period of risk (MPOR) is defined as nine business days plus the frequency of collateral calling, so daily calls will give a 10 business day MPOR.

Figure \ref{f:problemIM} shows the IM challenge to PFE: with IM, PFE(99\%) appears identically zero, ignoring for the moment the issue of exposure spikes from return of collateral (see next section). This is true even before we consider that IM is defined as the 99th percentile {\it calibrated including a period of stress}. In the figure, given a 20\%\ stress, the quantile for PFE would have to be defined as something above 99.86\%. Even if this re-definition of PFE was done, using such a high percentile for non-IM or non-collateralized counterparties would be problematic because it would be so high. The numbers that Credit Officers would be required to sanction would be completely outside previous experience. 

The problem for PFE with IM is not simply that PFE can be zero, it is that PFE is zero {\it and} we know that it ignores losses above its reference percentile (e.g. 95\%). This is a very uncomfortable situation: are the losses above 95\%\ small enough to ignore or not?

It may be argued that the collateral eligibility for IM is sufficiently wide to make the IM worthless. However, regulations are written specifically to avoid this including mark-to-market, haircuts, quality floors, etc.

Despite collateral and regulatory IM there can still be significant, if brief, exposure from spikes in exposure profiles due to return of collateral or similar. This is addressed by another recent market development, covered in the next section.

\subsection{Netting of mark-to-market flows and trade termsheet flows}

With collateralized counterparties spikes in exposure are observed on coupon and principle payment dates when collateral and termsheet flows are not netted. These spikes are from failure to return collateral following a termsheet payment. \cite{andersen2017rethinking} have pointed out that these spikes may mean that regulatory IM does not reduce exposure by 99\%, but perhaps only by 90\%\ in some cases.

Market services are now appearing that net collateral and termsheet flows. It is not clear whether they will eliminate spikes in default situations but if they do the, addition of IM will produce effectively zero exposure below the 99th percentile. This renders PFE(95\%), PFE(97.5\%), and PFE(99\%) of questionable utility for these counterparties.

\section{Potential Future Loss}

Given the existing and the new challenges to PFE for counterparty trading limits described above, we now introduce Potential Future Loss (PFL), Adjusted PFL, and Protected Adjusted PFL to address them.

\vskip3mm\noindent
\begin{define}[$\PFL_\M(t,q)$] Potential Future Loss at time $t$ in the future for quantile $q$ under measure \M\ is the future profile of Expected Shortfall(q) times Loss Given Default, i.e.
\begin{align}
\PFL_\M(t,q) :=& \E_\M\left[\lgd(t) \times V(\Pi,t,\delta_B,\delta_C) \vphantom{\Big|} \right. \nonumber \\ 
&\quad\ \ \ \left. \Big| {\lgd(t) \times V(\Pi,t,\delta_B,\delta_C) \ge b} \right]\\
b :=& \CDF_\M^{-1}(q)\left(\max(\lgd(t) \times V(\Pi,t,\delta_B,\delta_C), 0)\vphantom{\int}\right) 
\end{align}
\end{define}
\noindent
Notation as for \PFE\ in Table \ref{t:notation}. The \LGD\ is inside the expectation to take into account potential correlation between exposure $V$ and LGD. We expect that with the emphasis in FRTB-CVA \citep{bcbs2017finishing} on WWR modelling, that WWR will be widely implemented in that timescale. WWR includes changes in exposure with LGD as well as changes in exposure with credit quality \citep{green2016xva}. Exposure and LGD can be linked via correlation of exposure with credit quality, and correlation between credit quality and LGD \citep{altman2005link,frye2013loss}. Our definition of \PFL\ includes these aspects naturally as \PFL\ includes LGD.

If we were to assume that portfolio value and LGD were independent then
\begin{align}
\PFL_\M(t,q) :=& \E_\M[\lgd(t)] \times \E_\M\left[ V(\Pi,t,\delta_B,\delta_C) | {V(\Pi,t,\delta_B,\delta_C) \ge b} \right]\\
b :=& \CDF_\M^{-1}(q)\left(\max( V(\Pi,t,\delta_B,\delta_C), 0)\vphantom{\int}\right) 
\end{align}

\subsection{Adjusted PFL (aPFL)}

This extension to \PFL\ deals with the overlap with CVA. Both the limit for PFL and the profile calculation for PFL are changed.

{\bf Motivation for Adjusted PFL (aPFL):} PFL give a profile of future potential losses and a PFL limit gives a limit on potential future losses. Now incurred CVA is a loss that has already gone through PnL so it is not reasonable to ignore this when considering a limit on future losses. Ignoring incurred CVA is saying that future losses should ignore existing losses as though they had not happened. Hence we propose aPFL to incorporate incurred CVA as a flat constant negative shift on the PFL {\it limit} and a flat constant negative shift on the PFL {\it profile}.

It may appear that it is optional as to whether to subtract incurred CVA from PFE limits, when subtracting incurred CVA from losses. That is, it may appear that this is just a local policy choice. However this is not the case. Consider a counterparty that is getting progressively worse. If incurred CVA is only subtracted from the loss, as incurred CVA increases then the trading capacity will also increase. This is undesireable behaviour. If we subtract from both the limit and the PFE then we capture both effects. 

It may appear that by subtracting incurred CVA from {\it both} the PFL limit and from the PFL profile we have not achieved anything. This is not correct because the effect on the PFL profile is non-linear: only paths which still have positive exposure will contribute to the new aPFL profile. In addition monitoring of PFL limit changes enables Risk to observe losses already taken in PnL by the Front Office and so foster coherent management of risk across front and middle office. \cite{pykhtin2011counterparty} discussed the interaction of incurred CVA with limits.

\vskip3mm\noindent
{\bf Adjusted Potential Future Loss(q), aPFL(q)} is the future profile of PFL(q) with incurred CVA removed, and where the associated limit has had incurred CVA removed.
\vskip3mm\noindent
\begin{define}[$\aPFL_\M(t,q)$] Adjusted Potential Future Loss at time $t$ in the future for quantile $q$ under measure \M\ is the future profile of Expected Shortfall(q) times Loss Given Defaul,t adjusted for incurred CVA $X$, i.e.
\begin{align}
\aPFL_\M(t,q) :=& \E_\M\left[(\lgd(t) \times V(\Pi,t,\delta_B,\delta_C) - X) \vphantom{\Big|} \right. \nonumber \\
& \quad\ \ \ \left. \Big| {\lgd(t) \times V(\Pi,t,\delta_B,\delta_C) - X \ge b} \right]\\
b :=& \CDF_\M^{-1}(q)\left(\max(\lgd(t) \times V(\Pi,t,\delta_B,\delta_C) - X, 0)\vphantom{\int}\right) 
\end{align}
\end{define}
An example is in the Numerical Examples Section later. 

This definition, and the one below, could be modified by instead subtracting the time zero expected forward CVA, i.e. without resimulation. This would avoid applying more and more ``expired'' CVA later in the profiles.

\subsection{Protected Adjusted PFL (paPFL)}

This extension to \PFL\ deals with credit mitigation, as well as incurred CVA.

{\bf Motivation for Protected Adjusted PFL (paPFL):} a CVA desk may hedge the credit risk of a counterparty. It seems unreasonable not to include this credit hedge, hence we propose adjusting the PFL {\it profile} to include the effect of the credit mitigation. We do not propose changing the PFL {\it limit} because credit mitigation does not make the bank willing to lose more. Instead, credit mitigation reduces losses.

We do not propose including future hedging actions. We propose including the mitigation of existing positions only. This is standard from a risk point of view for the following reasons. Although there may be a hedging strategy and even a hedging policy circumstances change. Giving credit for future actions is problematic from a Credit Officer point of view: how can a Credit Officer be sure that the actions will be carried out, and even that the market will permit them to be carried out? Typically if a credit crisis is bad enough (2008 Financial Crisis or later Greek Crisis) then the CDS market closes for the worst names, and the market may jump for others.

\vskip3mm\noindent
{\bf Protected Adjusted Potential Future Loss(q), paPFL(q)} is the future profile of PFL(q) with incurred CVA removed and existing credit protection, $Y(t)$, is included. The associated PFL limit has had incurred CVA removed but is not affected by existing credit protection.
\vskip3mm\noindent
\begin{define}[$\paPFL_\M(t,q)$] Protected Adjusted Potential Future Loss at time $t$ in the future for quantile $q$ under measure \M\ is the future profile of Expected Shortfall(q) times Loss Given Default, adjusted for incurred CVA $X$ and existing credit protection, $Y(t)$, that directly references the counterparty, i.e.
\begin{align}
\paPFL_\M(t,q) :=& \E_\M\left[(\lgd(t) \times V(\Pi,t,\delta_B,\delta_C) - X - Y(t)) \vphantom{\Big|} \right. \nonumber\\
&\left. \quad\ \ \ \Big| {\lgd(t) \times V(\Pi,t,\delta_B,\delta_C) - X - Y(t) \ge b} \right]\\
b :=& \CDF_\M^{-1}(q)\left(\max(\lgd(t) \times V(\Pi,t,\delta_B,\delta_C) - X - Y(t), 0)\vphantom{\int}\right) 
\end{align}
\end{define}

Credit mitigation from a CDS is flat up to maturity of the CDS with a value of LGD times CDS notional. We do not recommend using the Regulatory approach to credit mitigation in counterparty credit risk because this only changes default probability. This is inconsistent with the concept of potential future loss which assumes that default has occurred, so changes in default probability are not relevant. The regulatory approach would change incurred CVA but we see this as secondary because the focus is on losses assuming default.

\subsection{Limit-setting Process}

Because PFL is comparable across counterparties and within counterparties by design the limit setting process can be much more systematic and transparent. First an extreme loss appetite can be set as the bank's risk appetite for derivatives. We call this extreme because it is not an expected loss but a high percentile. The bank can then apportion this extreme loss appetite to different sectors and counterparties according to the competitive advantage and business opportunities. As opportunities change the appetite can be re-apportioned transparently: a given amount of PFL limit in one place is comparable to a given amount of PFL limit in any other place. Executives can view the PFE limits and their usage at any granularity w.r.t. counterparties and this will be meaningful. With PFE this simplicity consistency of risk control is simply impossible.

\subsection{Recovery Rates}

There are no liquid instruments providing market implied recovery rates. However, many industry studies exist on sector-wide recovery rates and their variation with market stress \citep{dullmann2004systematic,altman2005link,frye2013loss}. Seniority-dependent recovery rate observations are also available \citep{jankowitsch2014determinants}. Beyond this bank Know Your Customer (KYC), Relationship Mangers, and Credit Officers together with internal (real-world) risk models, and market data service providers give inputs to internally computed recovery rates for use in PFL.

\section{Numerical Examples}

We demonstrate PFE and PFL (with variants) for a vanilla 10 year ATM USD IRS. The interest rate dynamics use a CIR stochastic volatility Libor Market Model calibrated to coterminal swaptions and the 5x5 swaption smile as in \citep{green2017bdy}. 

\subsubsection{Uncollateralized}

\begin{figure}[htbp]
\centering
\includegraphics[width=0.49\textwidth,clip,trim=0 0 0 0]{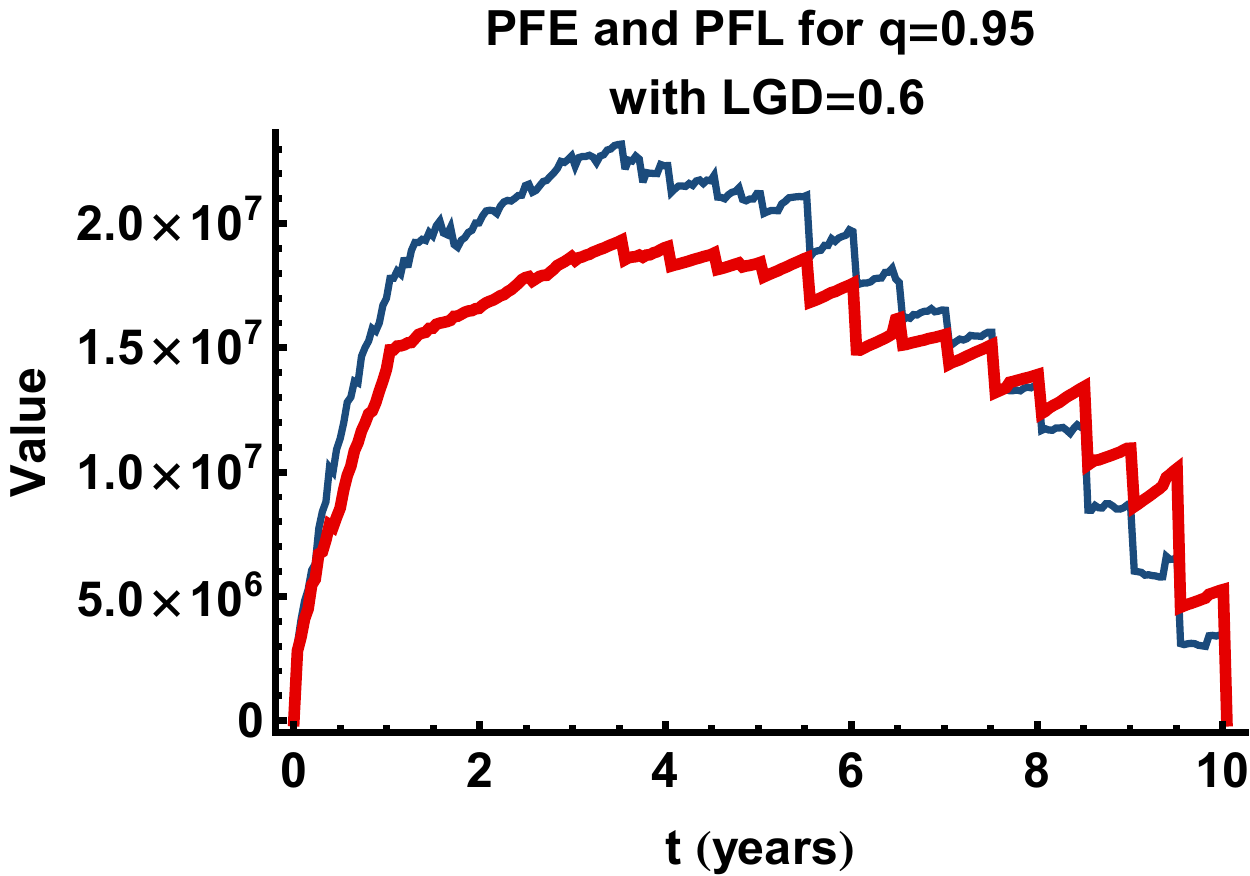}
\includegraphics[width=0.49\textwidth,clip,trim=0 0 0 0]{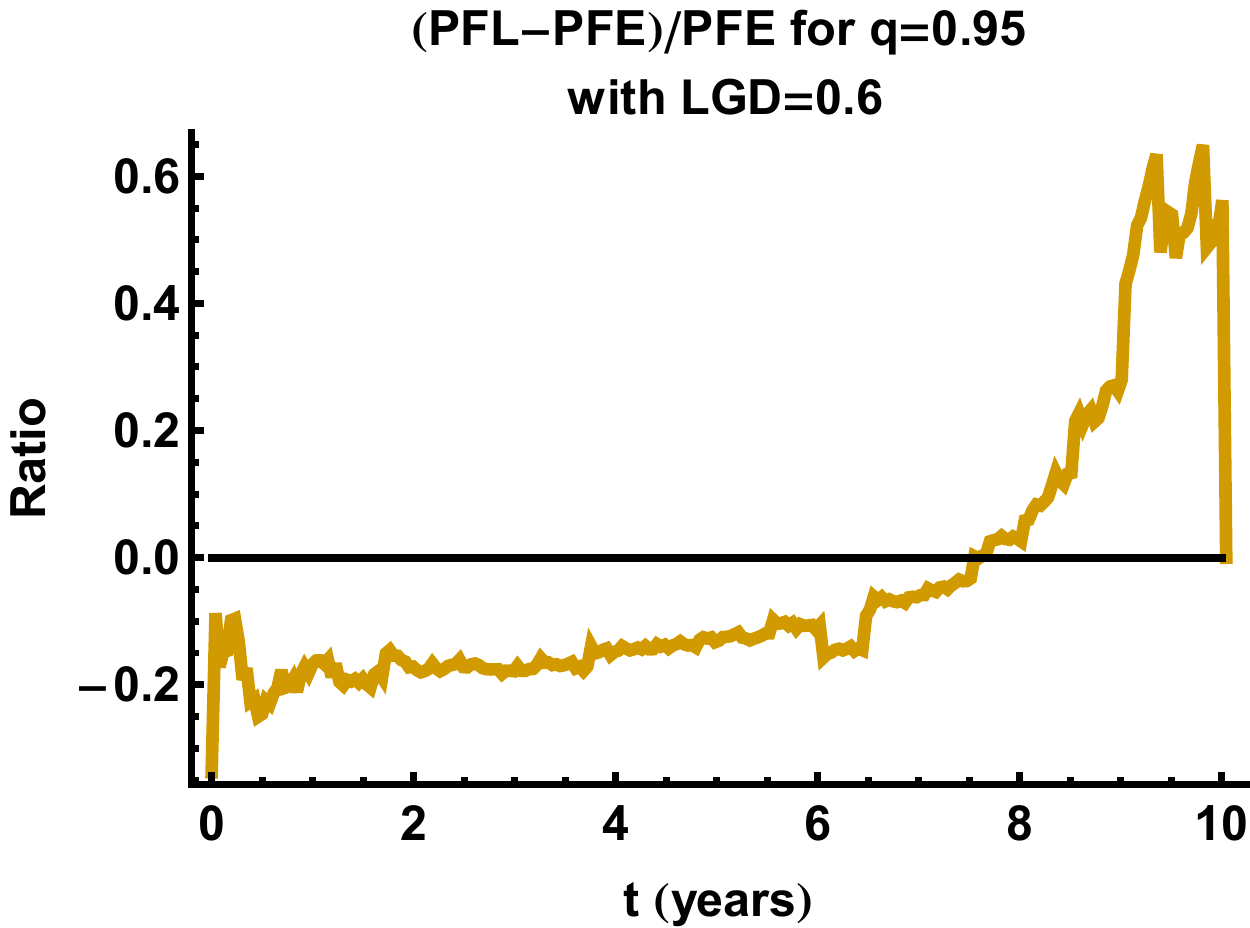}
\caption{LEFT: PFE (blue) and PFL (red, thicker) for uncollateralized 10 year ATM USD IRS with notional 100M. RIGHT: Comparison of PFL to PFE. The change in difference over the lifetime of the IRS comes from the change in exposure distribution over the lifetime of the IRS.}
\label{f:usd}
\end{figure}

Figure \ref{f:usd} compares PFL and PFE for the uncollateralized IRS over its lifetime. It is striking that the PFL is roughly similar to PFE when we recall that the PFL incorporates a 60\%\ LGD. This indicates that PFE is ignoring a very significant exposure tail above the 95\%\ quantile, even for such an ordinary product.

The RIGHT plot in Figure \ref{f:usd} gives the ratio (PFL-PFE)/PFE. The change in this ratio over the lifetime of the IRS indicates the change in the exposure distribution above the 95\%\ quantile. Not only does a VaR-type measure ignore this but the change in ratio cannot be captured with a simple multiplier because the ratio changes so much: from ${}-0.2$ to ${}+0.6$.

\begin{figure}[htbp]
\centering
\includegraphics[width=0.49\textwidth,clip,trim=0 0 0 0]{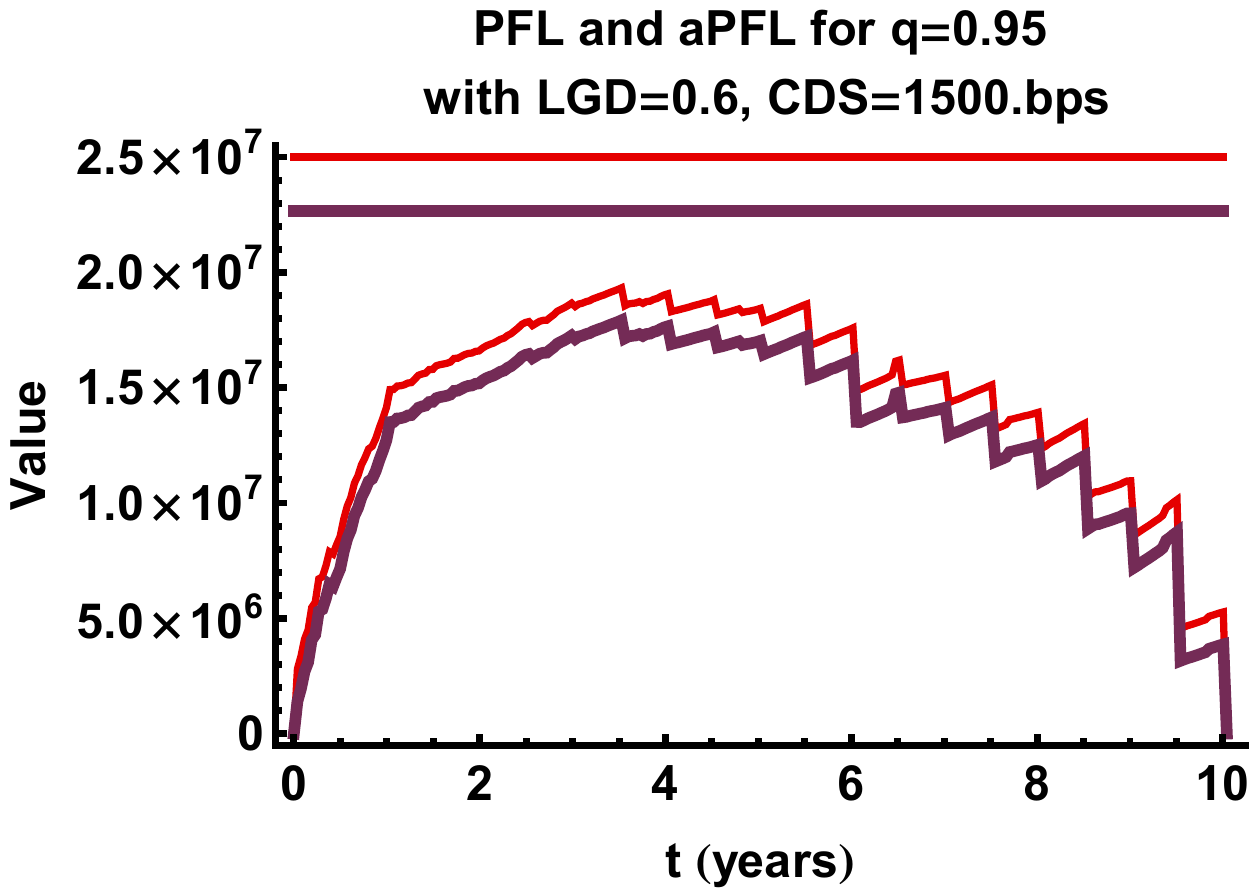}
\includegraphics[width=0.49\textwidth,clip,trim=0 0 0 0]{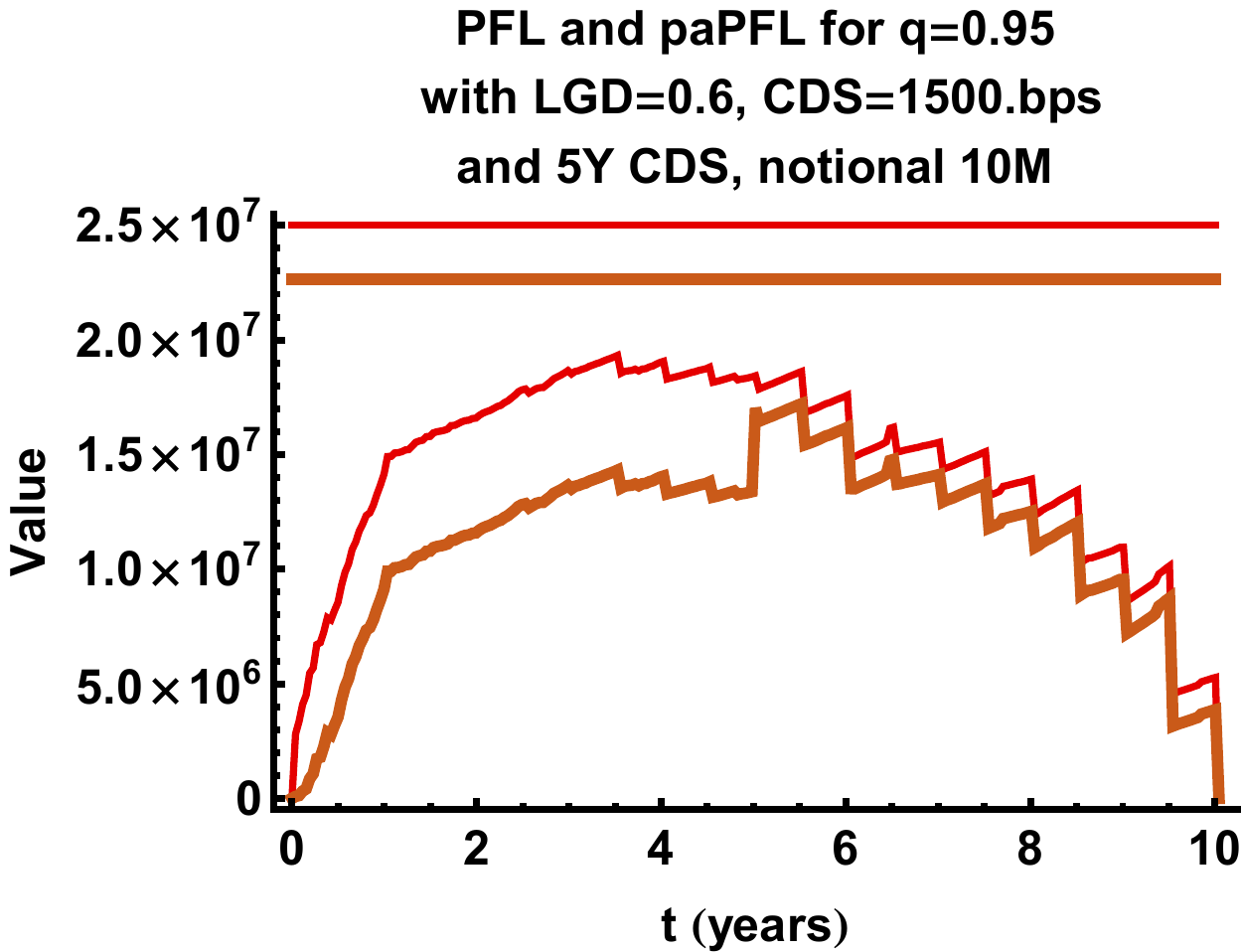}
\caption{LEFT: PFL (red, thin) and Adjusted PFL (plum, thicker) for uncollateralized 10 year ATM USD IRS with notional 100M and CDS spread 1500bps. RIGHT: PFL (red, thin) and Protected Adjusted PFL (orange, thicker) where there is now bought protection from a 5Y CDS.}
\label{f:usdAP}
\end{figure}

The LEFT plot in Figure \ref{f:usdAP} shows PFL and Adjusted PFL where the counterparty has a CDS spread of 1500bps which is probably the highest commonly observable before default. The incurred CVA has been subtracted from the PFL limit and the PFL profile to get aPFL. The non-linear effect of subtracting incurred CVA is clear as the difference between the limits is greater than the difference between the PFL and aPFL profiles.

The RIGHT panel in Figure \ref{f:usdAP} shows PFL and Protected Adjusted PFL in the case where a 5Y CDS with notional 10M has been purchased. The maximum expected positive exposure is under 6M (data not shown) so from an expectation point of view the CDS may remove all exposure up to 5Y. However, the range of exposures goes much higher than 6M so the effect of 10M of CDS for 5Y (assuming an LGD of 0.6) is much less than might be hoped. Thus we observe again how the non-linearity of exposure and the distribution of exposure combine to produce risk that is highly expensive to remove. A contingent CDS would remove all the exposure but these are bespoke and their sellers are familiar with the observations in this section.

\subsubsection{Collateralized, with IM, and with Flow Netting}

We now consider the 10Y IRS example with collateralization with an MPOR of 10 business days, zero minimum amount, zero threshold, and daily exchanges in USD cash.

\begin{figure}[htbp]
\centering
\includegraphics[width=0.49\textwidth,clip,trim=0 0 0 0]{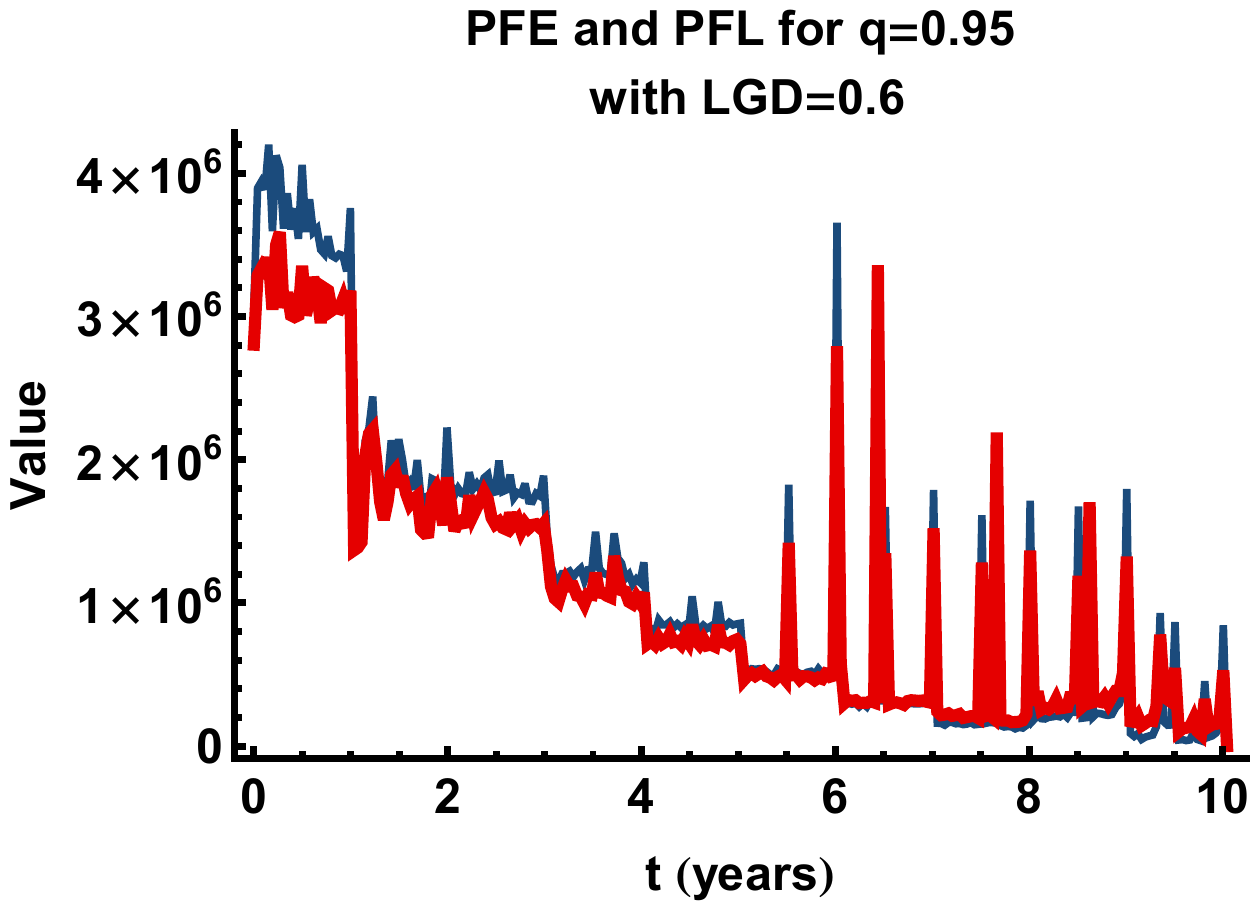}
\includegraphics[width=0.49\textwidth,clip,trim=0 0 0 0]{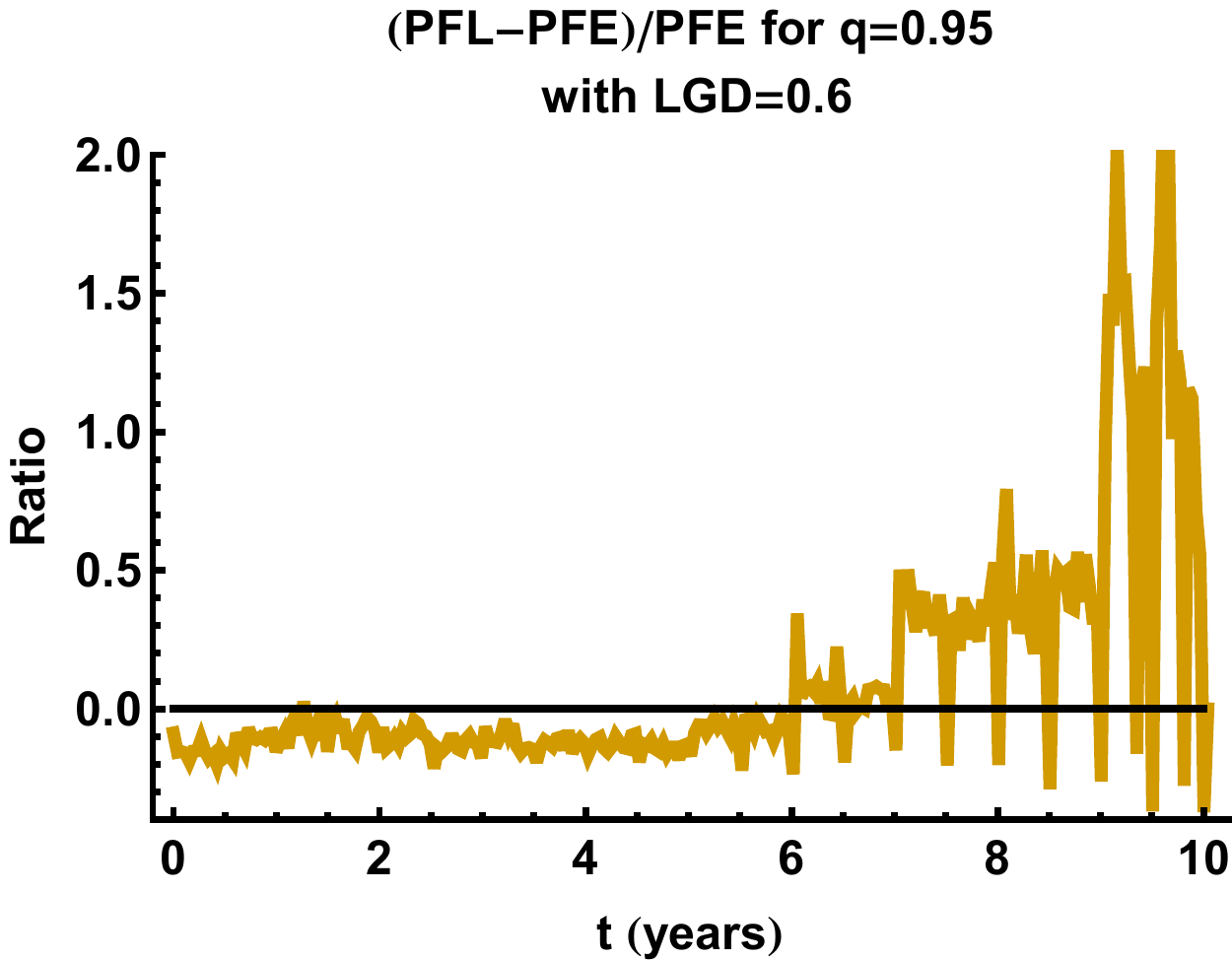}
\caption{PFE (blue) and PFL (red, thicker) for collateralized 10 year ATM USD IRS with notional 100M. RIGHT: Comparison of PFL to PFE. The change in difference over the lifetime of the IRS comes from the change in exposure distribution over the lifetime of the IRS and is more extreme because of collateralization.}
\label{f:usdCollat}
\end{figure}

\begin{figure}[htbp]
\centering
\includegraphics[width=0.49\textwidth,clip,trim=0 0 0 0]{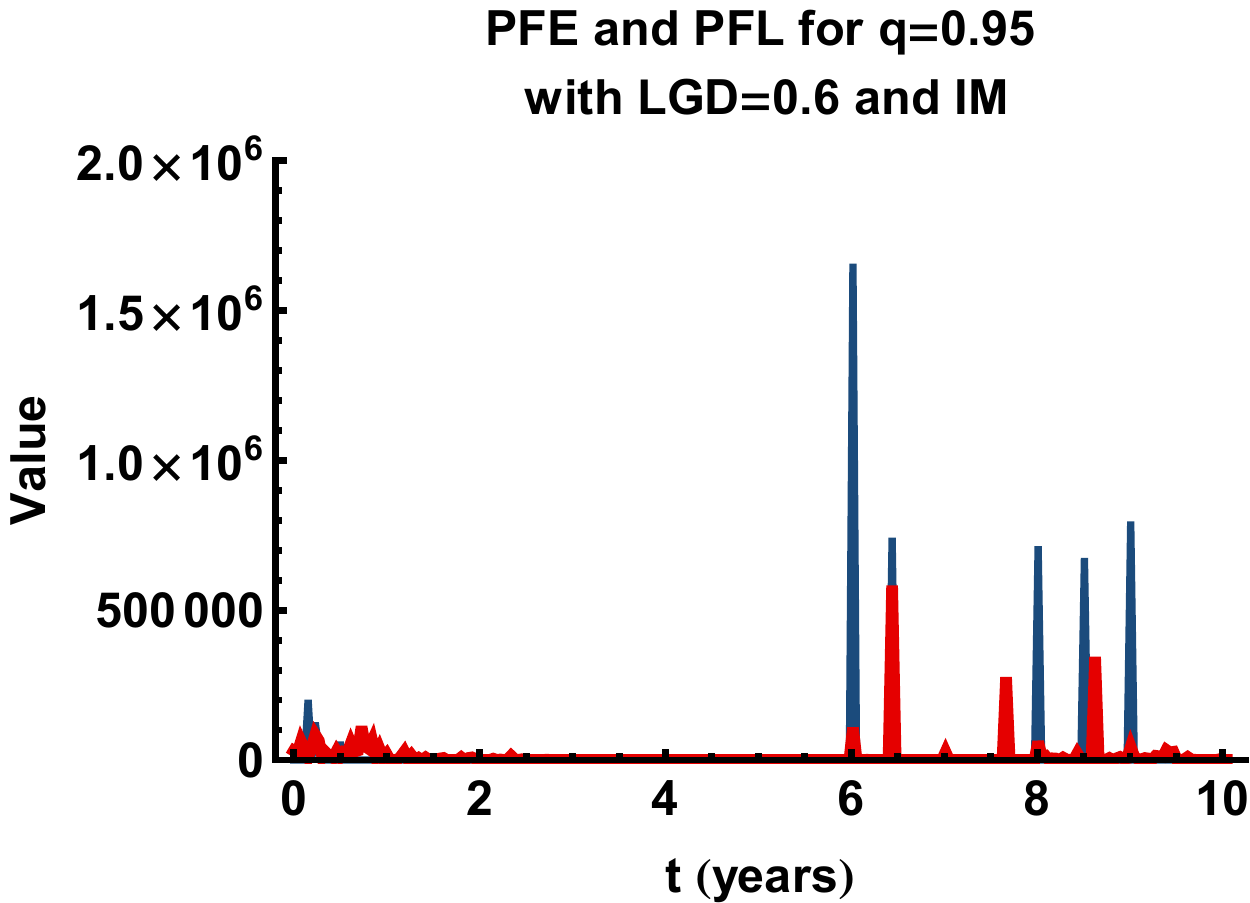}
\includegraphics[width=0.49\textwidth,clip,trim=0 0 0 0]{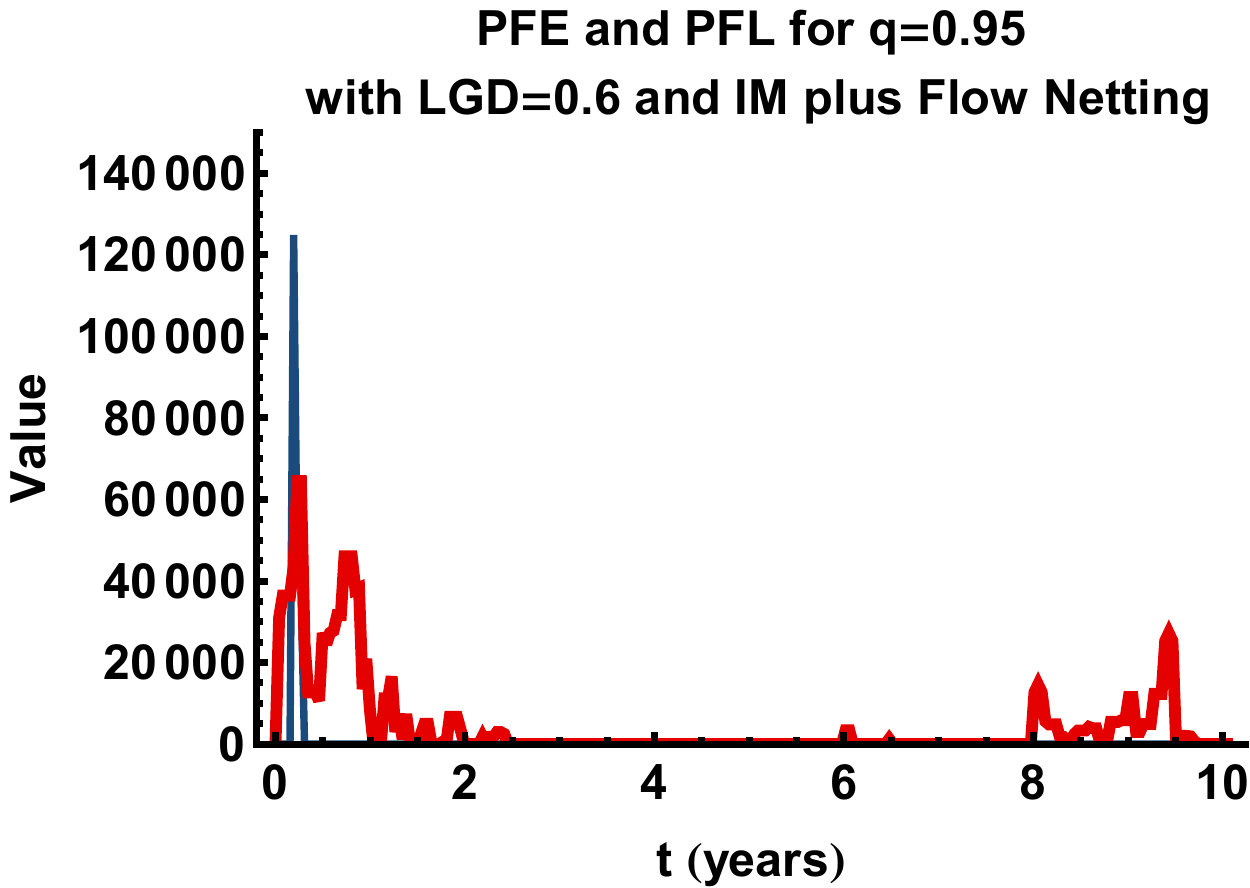}
\caption{LEFT: PFE (blue) and PFL (red, thicker) for collateralized 10 year ATM USD IRS with notional 100M with Schedule-based IM. RIGHT: As LEFT but with netting between trade and MtM flows. Note the difference in vertical scales.}
\label{f:usdCollatIMFlow}
\end{figure}

The LEFT plot in Figure \ref{f:usdCollat} shows the PFE and PFL profiles which are again roughly comparable despite PFL being calculated with an LGD of 0.6. 
The RIGHT plot shows the ratio of (PFL-PFE)/PFE and we see that there are highly significant differences (ratio of 0.5 and above) after 7 years. This is more extreme than in the uncollateralized case because the collateralization also changes the distribution above the reference percentile. Thus PFE is ignoring more of a distribution issue with collateralised counterparties than with uncollateralized counterparties.

The LEFT plot in Figure \ref{f:usdCollatIMFlow} shows the effect of Schedule-based IM. The exposure spikes from return of collateral are well known. When we consider the RIGHT plot where there is additionally netting between MtM flows and trade flows we see that even for PFL there are considerable stretches where there is effectively zero PFL. {\it With PFL} this is useful information because we know that there is no ignored exposure above the chosen percentile. Where PFL is effectively zero then there is effectively zero credit risk, and we can be certain of this. There may be liquidity risk but that is not counterparty credit risk.

\section{Conclusions}

Developing challenges to PFE in terms of widespread IM and netting of collateral and trade flows mean that PFE will become of questionable value (identically zero, but ignoring losses above its reference percentile) as a counterparty trading limit. Outside of widespread IM and netting of collateral and trade flows, pre-existing challenges to PFE (comparability across counterparties, exposure distribution shapes, collateralization, multiple seniorities, ignoring existing credit losses, ignoring credit mitigation) mean that it is already a poor fit for purpose. We propose using expected shortfall times loss given default to arrive at Potential Future Loss (PFL). PFL, together with Adjusted and Protected versions (including incurred CVA, and credit protection respectively) are robust against both pre-existing challenges and developing challenges to PFE.

\section*{Acknowledgements}

The authors would gratefully like to acknowledge feedback from participants at the MVA Round-table (September 2017, Canary Wharf), QuantMinds (May 2018, Lisbon) and discussions with Sebastian Steinfeld and Helmut Glemser.

\bibliographystyle{chicago}
\bibliography{XVAbibliography}

\end{document}